\RequirePackage{ifpdf}
\ifpdf 
\documentclass[pdftex]{sigma}
\else
\documentclass{sigma}
\fi

\numberwithin{equation}{section}

\begin{document}


\renewcommand{\PaperNumber}{023}

\FirstPageHeading

\renewcommand{\thefootnote}{$\star$}

\ShortArticleName{SUSY Quantum Hall Ef\/fect on
Non-Anti-Commutative Geometry}

\ArticleName{SUSY Quantum Hall Ef\/fect\\ on Non-Anti-Commutative
Geometry\footnote{This paper is a contribution to the Proceedings
of the Seventh International Conference ``Symmetry in Nonlinear
Mathematical Physics'' (June 24--30, 2007, Kyiv, Ukraine). The
full collection is available at
\href{http://www.emis.de/journals/SIGMA/symmetry2007.html}{http://www.emis.de/journals/SIGMA/symmetry2007.html}}}

\Author{Kazuki HASEBE}

\AuthorNameForHeading{K. Hasebe}

\Address{Department of General Education, Takuma National College of Technology,\\
 Takuma-cho, Mitoyo-city, Kagawa 769-1192, Japan}

\Email{\href{mailto:hasebe@dg.takuma-ct.ac.jp}{hasebe@dg.takuma-ct.ac.jp}}

\ArticleDates{Received October 01, 2007, in f\/inal form February
07, 2008; Published online February 22, 2008}

\Abstract{We review the recent developments of the SUSY quantum
Hall ef\/fect
[\mbox{\href{http://arxiv.org/abs/hep-th/0409230}{hep-th/0409230}},
\href{http://arxiv.org/abs/hep-th/0411137}{hep-th/0411137},
\href{http://arxiv.org/abs/hep-th/0503162}{hep-th/0503162},
\href{http://arxiv.org/abs/hep-th/0606007}{hep-th/0606007},
\href{http://arxiv.org/abs/0705.4527}{arXiv:0705.4527}]. We
introduce a SUSY formulation of the quantum Hall ef\/fect on
supermanifolds.
 On each of supersphere and superplane, we investigate  SUSY Landau problem and explicitly construct
 SUSY extensions of  Laughlin wavefunction and topological excitations.
The non-anti-commutative geometry naturally emerges in the lowest
Landau level and  brings particular physics to the SUSY quantum
Hall ef\/fect. It is shown that   SUSY provides a unif\/ied
picture of the original Laughlin and Moore--Read states. Based on
the charge-f\/lux duality, we also develop  a  Chern--Simons
ef\/fective f\/ield theory for the SUSY quantum Hall ef\/fect.}


\Classification{17B70; 58B34; 81V70}

\renewcommand{\thefootnote}{\arabic{footnote}}
\setcounter{footnote}{0}

\section{Introduction}

Quantum Hall ef\/fect (QHE) provides a rare physical set-up for
the noncommutative geometry (NCG), where  the center-of-mass
coordinates of electron satisfy the NC algebra
\begin{gather*}
[X,Y]=i\ell_B^2.
\end{gather*}
Phenomena observed in QHE  are governed by NCG
 and manifest its peculiar  properties \cite{hep-th/0209198}.
Until recently, it was believed  QHE could be formulated only in
2D space. However, a few years ago, a 4D generalization of the QHE
was successfully formulated  in~\cite{cond-mat/0110572}. The 4D
QHE exhibits  reasonable analogous physics  observed in 2D QHE,
such as  incompressible quantum liquid, fractionally charged
excitations, massless edge modes and etc. The appearance of the 4D
QHE was a breakthrough for  sequent  innovational progress of
generalizations of the QHE. By many authors, the formulation of
QHE has been quickly extended  on  various higher dimensional
manifolds, such as  complex projected spaces
\cite{hep-th/0203264},  fuzzy spheres \cite{cond-mat/0306045,
hep-th/0310274,hep-th/0309212},  Bergman ball
\cite{hep-th/0505095},  a~f\/lag manifold $F_2$
\cite{hep-th/0610157} and $\theta$-deformed manifolds
\cite{hep-th/0504092}. The  developments of  QHE  have attracted
many attentions from   non-commutative geometry and matrix model
researchers, since   higher dimensional structures of NCG  are
physically realized in the set-up of the higher dimensional QHE.
Indeed, the analyses of the higher dimensional QHE  have provided
deeper  understandings of  physical properties of  NCG and matrix
models\footnote{For the higher dimensional developments of QHE and
relations to  fuzzy geometries and matrix models, interested
readers may consult~\cite{hep-th/0606161} as  a good review.}.
Besides,  3D reduction of the 4D QHE gave a clue for the
theoretical discovery of  the spin Hall ef\/fect
\cite{cond-mat/0308167} in condensed matter physics.

In this review, we report a new  extension of  QHE:  a SUSY
extension of the QHE,  where particle carries fermionic center-of-mass degrees of freedom as well as bosonic ones: $(X,Y)$ and
$(\Theta_1,\Theta_2)$. They satisfy the SUSY NC relations
(non-commutative and non-anti-commutative   relations):
\begin{gather*}
[X,Y]=i\ell_B^2,\qquad \{\Theta_1,\Theta_2\}=\ell_B^2.
\end{gather*}
There are much  motivations to explore the SUSY  QHE. With the
developments of
 string theory,
it was found that the non-anti-commutative geometry  is naturally
 realized on D-brane in gravi-photon background
\cite{hep-th/0302078,hep-th/0302109, hep-th/0305248}. The
supermatrix models are  constructed based on the super Lie group
symmetries, and the  non-anti-commutative geometry is embedded in
supermatrix models by nature  \cite{hep-th/0102168,
hep-th/0311005}. The SUSY QHE would provide  a ``physical'' set-up
which such string theory related models attempt to describe, and
exhibit
 exotic features of the non-anti-commutative geometry in a most obvious way.
Apart from possible  applications to  string theory,
  construction of the  SUSY QHE  contains many interesting subjects of its own right.  QHE is deeply related to  exotic mathematical and physical ideas:
 fuzzy geometry, Landau problem,  Hopf f\/ibration  and topological f\/ield theory.
As we supersymmetrize  QHE,  we inevitably encounter these
structures. It is quite challenging  to extend them in
self-consistent SUSY frameworks, and interesting  to see how they
work.

Here, we mention  several   developments related to  SUSY QHE and
SUSY Landau problem. Algebraic and topological structures of the
fuzzy supersphere are well examined
in~\cite{hep-th/0204170,arXiv:hep-th/0506037}. Field theory models
on  supersphere have  already been proposed;    a non-linear sigma
model  and a scalar f\/ield model were explored
in~\cite{arXiv:hep-th/0311031} and~\cite{hep-th/0409257},
respectively. Numerical calculations on fuzzy spaces have also
been carried out (see~\cite{arXiv:hep-th/0609205} as a review).
The SUSY Landau problems on higher dimensional coset
supermanifolds were developed
in~\cite{hep-th/0311159,hep-th/0404108}. Specif\/ically
in~\cite{hep-th/0311159},   the fuzzy super geometry on
 complex projective superspace  $\mathbb{C}P^{n|m}=SU(n+1|m)/U(n|m)$ was explored in detail.
 Planar SUSY Landau models  were constructed in~\cite{ hep-th/0510019, hep-th/0612300} where  the negative norm problem as well as  its cure were discussed. (These works have several overlaps with our planar SUSY Landau problem developed in Subsection~\ref{subsecplanarlandaupro}.)
The spherical SUSY Landau problem with $\mathcal{N}=4$ SUSY was
also investigated in~\cite{hep-th/0503244}.  Embedding of  SUSY
structure to QH matrix model has been explored
in~\cite{hep-th/0410070}. More recently, SUSY-based analysis was
applied to edge excitations on the $5/2$ f\/illing QH
state~\cite{0706.1338}.

\section{Preliminaries}

A nice set-up for the QHE, needless to consider  boundary
ef\/fects, is given by  Haldane \cite{HaldanePRL51}  who
formulated QHE on two-sphere  with
 Dirac monopole at its center.
 We supersymmetrize Haldane's system by replacing  bosonic sphere  with supersphere, and Dirac monopole with supermonopole (see Table~\ref{SUSYextensionofHaldane}).
\begin{table}
\renewcommand{\arraystretch}{1.5}
\centering \caption{The original Haldane's set-up and our SUSY
extension.}\vspace{1mm}
\begin{tabular}{|c|c|c|}
 \hline    & The original Haldane's set-up & ~~~~~~~~ Our SUSY set-up ~~~~~~~  
\\
 \hline
Base manifold &  $S^{2}=SU(2)/U(1)$ &  $S^{2|2}=OSp(1|2)/U(1)$
\\ \hline
Monopole & Dirac monopole  &   Supermonopole
\\
\hline
  Hopf map & $S^3 \rightarrow S^2$ & $S^{3|2} \rightarrow S^{2|2}$ \\
\hline
Emergent fuzzy manifold & Fuzzy sphere & Fuzzy supersphere \\
\hline Many-body groundstate  & $SU(2)$ invariant Laughlin
& $OSp(1|2)$ invariant Laughlin \\
\hline
\end{tabular}\label{SUSYextensionofHaldane}
\end{table}
The supersphere $S^{2|2}$ is a coset manifold taking the form of
$OSp(1|2)/U(1)$. The adaptation of  coset supermanifold has an
advantage  that the SUSY is automatically embedded by the coset
construction.
 Here, we introduce basic  mathematics  needed to explore SUSY QHE.

\subsection[The $OSp(1|2)$ super Lie algebra {[30,35]}]{The $\boldsymbol{OSp(1|2)}$ super Lie algebra \cite{booksuperlie,PRL94206802}}

 The $OSp(1|2)$ group is a super Lie group whose bosonic generators $L_a$ ($a=x,y,z$) and fermionic generators $L_{\alpha}$ ($\alpha=\theta_1,\theta_2$)
satisfy the SUSY algebra:
\begin{gather}
[L_a,L_b]=i\epsilon_{abc}L_c,\qquad
[L_a,L_{\alpha}]=\tfrac{1}{2}(\sigma_a)_{\beta\alpha}L_{\beta},\qquad
\{L_{\alpha},L_{\beta}\}=\tfrac{1}{2}(C\sigma_a)_{\alpha\beta}L_a,
\label{ospalgebrarelation}
\end{gather}
where $C$ denotes the charge conjugation matrix for $SU(2)$ group,
\begin{gather}
C=
\begin{pmatrix}
0 & 1 \\
-1 & 0
\end{pmatrix}.\label{chargesu2mat}
\end{gather}
The $OSp(1|2)$ algebra contains the $SU(2)$ subalgebra, and  in
the $SU(2)$ language, $L_a$ are $SU(2)$ vector and $L_{\alpha}$
are  $SU(2)$ spinor. The dif\/ferential operators that satisfy the
$OSp(1|2)$ algebra~are
\begin{gather}
{M}_a=-i\epsilon_{abc}x_b\partial_c+\tfrac{1}{2}(\sigma_a)_{\alpha\beta}\partial_{\beta},\nonumber\\
{M}_{\alpha}=\tfrac{1}{2}(C\sigma_a)_{\alpha\beta}x_a\partial_{\beta}-
\tfrac{1}{2}\theta_{\beta}(\sigma_a)_{\beta\alpha}\partial_a.\label{freeospalgebra}
\end{gather}
The Casimir operator for the $OSp(1|2)$ group is given by
$L_a^2+C_{\alpha\beta}L_{\alpha}L_{\beta}$ whose eigenvalue is
   $S(S+\frac{1}{2})$ with integer or half-integer Casimir index $S$.
The dimension of the irreducible representation specif\/ied by the
Casimir
 index $S$ is $4S+1$. In terms of  $SU(2)$, the $OSp(1|2)$ irreducible representation  $4S+1$ is decomposed to
 $2S+1\oplus 2S,$
where $2S+1$ and $2S$  are $SU(2)$ irreducible representations
with $SU(2)$ Casimir index $S$ and $S-{1}/{2}$, respectively.
Since their $SU(2)$ spin quantum numbers are dif\/ferent by $1/2$,
they are regarded as SUSY partners. The $OSp(1|2)$ matrices for
the fundamental representation ($S=1/2$)  are given by
 the following  $3\times 3$  matrices:
\begin{gather*}
l_a=\tfrac{1}{2}
\begin{pmatrix}
\sigma_a & 0 \\
0 & 0
\end{pmatrix},\qquad
l_{\alpha}=\tfrac{1}{2}
\begin{pmatrix}
0 & \tau_{\alpha} \\
-(C\tau_{\alpha})^t & 0
\end{pmatrix},
\end{gather*}
where  $\sigma_a$ are Pauli matrices, $C$ is the
charge-conjugation matrix (\ref{chargesu2mat}),
 and $\tau_{1}=(1,0)^t$, $\tau_{2}=(0,1)^t$.
 These matrices are super-hermitian, in the sense,
 \begin{gather*}
 l_a^{\ddagger}=l_a, \qquad l_{\alpha}^{\ddagger}=C_{\alpha\beta}l_{\beta},
 \end{gather*}
where the super adjoint $\ddagger$ is def\/ined by
\begin{gather*}
\begin{pmatrix}
 A & B \\
 C & D
\end{pmatrix}^{\ddagger}
=
\begin{pmatrix}
 A^{\dagger} & C^{\dagger} \\
 -B^{\dagger} & D^\dagger
\end{pmatrix}.
\end{gather*}
The complex representation  matrices corresponding to $l_a$ and
$l_{\alpha}$ are  constructed as
\begin{gather}
\tilde{l}_a=-l_a^*,\qquad
\tilde{l}_{\alpha}=C_{\alpha\beta}l_{\beta}.
\label{complexrepospgene}
\end{gather}
Short calculation shows that they actually satisfy
 the $OSp(1|2)$ algebra (\ref{ospalgebrarelation}).
It is important to note that the complex and the original
representations are unitary equivalent,
\begin{gather*}
\tilde{l}_a= \mathcal{R}^{\dagger} l_a \mathcal{R}, \qquad
\tilde{l}_{\alpha}= \mathcal{R}^{\dagger} l_{\alpha} \mathcal{R},
\end{gather*}
where the unitary matrix $\mathcal{R}$ is given by
\begin{gather*}
\mathcal{R}=
\begin{pmatrix}
0 & 1 & 0 \\
-1 & 0 & 0 \\
0 & 0 & -1
\end{pmatrix}.
\end{gather*}
Properties of $\mathcal{R}$ are summarized as
\begin{gather*}
\mathcal{R}^t=\mathcal{R}^{\dagger}=\mathcal{R}^{\ddagger}=\mathcal{R}^{-1},\qquad
\mathcal{R}^2=(\mathcal{R}^t)^2=
\begin{pmatrix}
-1 & 0 & 0 \\
0 & -1 & 0 \\
0 & 0 & 1
\end{pmatrix}.
\end{gather*}
$\mathcal{R}$ plays crucial roles in construction of  SUSY
Laughlin wavefunction and  topological excitations as we shall see
in Subsection~\ref{SphSUSYLaugh}.

\subsection[SUSY Hopf map and supermonopole {[32,33,34]}]{SUSY Hopf map and supermonopole \cite{LMP3145,math-ph/9907020,NPB70994}}

As is well known, the mathematical background of the Dirac
monopole is given by
 the  (1st) Hopf map (see for instance~\cite{booknakahara}).
Similarly, the supermonopole\footnote{The supermonopole is usually
referred to the graded monopole in literatures.}  is  introduced
as a SUSY extension of the  Hopf bundle~\cite{LMP3145}. As the
Hopf map is  the mapping from $S^3$ to $S^2$, the SUSY Hopf map is
the mapping from  $S^{3|2}$ to $S^{2|2}$, which is explicitly
\begin{gather}
\psi=
\begin{pmatrix}
u\\
v\\
\eta
\end{pmatrix}
\rightarrow
\frac{1}{R}({{x}_a},{{\theta}_{\alpha}})=2\psi^{\ddagger}(l_a,l_{\alpha})\psi,
\label{superHopfmap}
\end{gather}
or, with  the complex representation,
\begin{gather}
\psi=
\begin{pmatrix}
u\\
v\\
\eta
\end{pmatrix}
\rightarrow \frac{1}{R}({{x}_a},{{\theta}_{\alpha}})=
-2\psi^{t}(\tilde{l}_a,\tilde{l}_{\alpha})\psi^*.
\label{superHopfmap2}
\end{gather}
Here, $u$ and $v$ form Grassmann even $SU(2)$ spinor, while $\eta$
is  a Grassmann odd $SU(2)$ singlet. The superadjoint $\ddagger$
is def\/ined by $\psi^{\ddagger}=(u^*,v^*,-\eta^*)$.\footnote{It
is noted that the symbol *  does not denote the conventional
complex conjugation but denotes   the pseudo-conjugation that acts
to the Grassmann odd variables as
$(\eta_1\eta_2)^*=\eta_1^*\eta_2^*$ and $(\eta^*)^*=-\eta$.
See~\cite{booksuperlie} for more details.} With the constraint
$\psi^{\ddagger}\psi=1$, $\psi$ is regarded as the coordinate on
$S^{3|2}$, and  $(x_a,\theta_{\alpha})$ given by
equation~(\ref{superHopfmap}) (or equation~(\ref{superHopfmap2}))
automatically satisfy the relation that def\/ines the supersphere
$S^{2|2}$ with radius $R$,
\begin{gather}
x_a^2+C_{\alpha\beta}\theta_{\alpha}\theta_{\beta}=R^2.
\label{r2supersphere}
\end{gather}
 $x_a$ and $\theta_{\alpha}$ represent
bosonic and fermionic coordinates of supersphere, respectively.
The SUSY Hopf spinor is simply a super coherent state;
\begin{gather*}
\frac{1}{R}l_a \psi\cdot
{x_a}+\frac{1}{R}C_{\alpha\beta}l_{\alpha}\psi\cdot
\theta_{\beta}=\frac{1}{2}\psi,
\end{gather*}
or
\begin{gather*}
\frac{1}{R}\tilde{l}_a \psi^*\cdot
x_a+\frac{1}{R}C_{\alpha\beta}\tilde{l}_{\alpha}\psi^*\cdot
\theta_{\beta}=-\frac{1}{2}\psi^*,
\end{gather*}
as suggested by equation~(\ref{superHopfmap})  or
equation~(\ref{superHopfmap2}). The super Hopf spinor is
explicitly represented~as
\begin{gather}
\psi =\frac{1}{\sqrt{2R^3(R+x_3)}}
\begin{pmatrix}
 (R+x_3)\bigl(R-\frac{1}{4(R+x_3)}\theta C\theta\bigr)\vspace{1mm}\\
 (x_1+ix_2)\bigl(R+\frac{1}{4(R+x_3)}\theta C\theta\bigr)\vspace{1mm}\\
 (R+{x_3})\theta_1+({x_1}+i{x_2})\theta_2
\end{pmatrix}\cdot e^{i\chi},
\label{psiexplicit}
\end{gather}
where the $U(1)$ phase  $e^{i\chi}$ geometrically corresponds to
  $S^1$-f\/ibre  on $S^{2|2}$, and  is canceled in the SUSY Hopf map (\ref{superHopfmap}) or (\ref{superHopfmap2}).
With the expression (\ref{psiexplicit}),  the supermonopole gauge
f\/ields are explicitly calculated by   the Berry phase formula:
\begin{gather}
-i\psi^{\ddagger}d\psi= dx_a A_a +d\theta_{\alpha} A_{\alpha}.
\label{SUSYBerryphaseform}
\end{gather}
The results are
\begin{gather}
A_a=\frac{I}{2R(R+x_3)}\epsilon_{ab3}{x_b}\left(1+\frac{2R+x_3}{2R^2(R+x_3)}\theta C\theta\right),\nonumber\\
A_{\alpha}=i\frac{I}{2R^3}(\sigma_a
C)_{\alpha\beta}x_a\theta_{\beta},\label{supermonopolegauge}
\end{gather}
with $I=1$. $A_a$ and $A_{\alpha}$  form a  super-vector multiplet
under the $OSp(1|2)$ transformation, and they would be interpreted
as photon and photino f\/ields, respectively. The f\/ield
strengths are def\/ined by
\begin{gather*}
F_{ab}=\partial_a A_b-\partial_b A_a, \qquad
F_{a\alpha}=\partial_a A_{\alpha}-\partial_{\alpha}A_a,\qquad
F_{\alpha\beta}=\partial_{\alpha}
A_{\beta}+\partial_{\beta}A_{\alpha},
\end{gather*}
 and  obtained as
\begin{gather*}
F_{ab}=-\frac{I}{2R^3}\epsilon_{abc}x_c\left(1+\frac{3}{2R^2}\theta
C\theta\right),\qquad
F_{a\alpha}=i\frac{I}{2R^3}(\sigma_b C)_{\alpha\beta}\theta_{\beta} \left(\delta_{ab}-\frac{3}{R^2}x_ax_b\right),\\
F_{\alpha\beta}=i\frac{I}{R^3}x_a(\sigma_a
C)_{\alpha\beta}\left(1+\frac{3}{2R^2}\theta C\theta\right).
\end{gather*}
 Here, $I/2$
takes integer or half-integer and denotes the quantized
supermonopole charge. The magnitude of the supermonopole  magnetic
f\/ields is given by
\begin{gather}
B=\frac{4\pi {I}/{2}}{4\pi R^2}=\frac{I}{2R^2}.
\label{magneticfieldonsphere}
\end{gather}

\section[The spherical SUSY quantum Hall effect]{The spherical SUSY quantum Hall ef\/fect}

\subsection[The spherical SUSY Landau problem {[34]}]{The spherical SUSY Landau problem \cite{NPB70994}}\label{hamiltoniansphericalSUSYLandau}

 With the above set-up, we discuss  one-particle problem on   a  supersphere in  a supermonopole background.
Since the particle on a supersphere is concerned,
 the Hamiltonian  does not contain the radial part, and
is simply given by  the angular part
\begin{gather}
H=\frac{1}{2MR^2}(\Lambda_a^2+C_{\alpha\beta}\Lambda_{\alpha}\Lambda_{\beta}),
\label{SUSYoneHam}
\end{gather}
where $\Lambda_a$ ($a=1,2,3)$ and $\Lambda_{\alpha}$
$(\alpha=1,2)$ represent the SUSY   covariant angular momenta
constructed from~(\ref{freeospalgebra}) with  replacing
the  partial derivatives to the covariant deriva\-tives:
\begin{gather}
\Lambda_{a}=-i\epsilon_{abc}x_b(\partial_c+iA_a)
+\tfrac{1}{2}\theta_{\alpha}(\sigma_a)_{\alpha\beta}(\partial_{\beta}+iA_{\beta}),\nonumber\\
\Lambda_{\alpha}=\tfrac{1}{2}(C\sigma_a)_{\alpha\beta}x_a(\partial_{\beta}+iA_{\beta})
-\tfrac{1}{2}\theta_{\beta}(\sigma_a)_{\beta\alpha}(\partial_a+iA_a).\label{lambdas}
\end{gather}
Here, $A_a$ and $A_{\alpha}$ are  the supermonopole gauge f\/ields
(\ref{supermonopolegauge}).
Since the motion of particle is  conf\/ined on the supersphere,
$\Lambda_a$ and $\Lambda_{\alpha}$ are tangent to the superface of
the supersphere:
\begin{gather}
\Lambda_a x_a+C_{\alpha\beta}\Lambda_a \theta_{\beta}=x_a
\Lambda_a+C_{\alpha\beta}\theta_{\alpha}\Lambda_{\beta}=0.
\label{SUSYorthogonalityLandx}
\end{gather}
The covariant momenta are not conserved quantities (for f\/inite
$A_{a}$ and $A_{\alpha}$), and  they do not  exactly satisfy the
$OSp(1|2)$ algebra:
\begin{gather}
[\Lambda_a,\Lambda_b]=i\epsilon_{abc}\left(\Lambda_c-\frac{I}{2R}x_c\right),\qquad
[\Lambda_a,\Lambda_{\alpha}]=\tfrac{1}{2}(\sigma_a)_{\beta\alpha}\left(\Lambda_{\beta}-\frac{I}{2R}\theta_{\beta}\right),
\nonumber\\
\{\Lambda_{\alpha},\Lambda_{\beta}\}=\tfrac{1}{2}(C\sigma_a)_{\alpha\beta}\left(\Lambda_a-\frac{I}{2R}x_a\right).
\label{lambdaalgebras}
\end{gather}
The ``extra'' terms in the right-hand-sides of
(\ref{lambdaalgebras}) are proportional to
 the supermonopole magnetic f\/ields,
\begin{gather*}
B_a=-\frac{I}{2R^3}x_a,\qquad
B_{\alpha}=-\frac{I}{2R^3}\theta_{\alpha}.
\end{gather*}
The magnetic f\/ield $B$ (\ref{magneticfieldonsphere}) is equal to
the magnitude
  of $B_a$ and $B_{\alpha}$: $B=\sqrt{B_a^2+C_{\alpha\beta}B_{\alpha}B_{\beta}}$. The number of the magnetic cells each of which occupies the area, $2\pi\ell_B^2$ with  magnetic length $\ell_B=1/\sqrt{B}$, on the supersphere is given by
\begin{gather*}
N_{\Phi}=\frac{4\pi R^2}{2\pi\ell_B^2}=  I.
\end{gather*}
Adding the angular momenta of the supermonopole f\/ields to
covariant angular momenta, the conserved $OSp(1|2)$ angular
momenta  are constructed as
\begin{gather}
L_a=\Lambda_a-\frac{I}{2R}x_a,\qquad
L_{\alpha}=\Lambda_{\alpha}-\frac{I}{2R}\theta_{\alpha}.
\label{llambdax}
\end{gather}
$L_a$ are   $SU(2)$ rotation generators, and $L_{\alpha}$ play the
role of  supercharges in this model. It is straightforward to
check that $L_a$ and $L_{\alpha}$ exactly satisfy  the $OSp(1|2)$
algebra (\ref{ospalgebrarelation}).

 From the orthogonality relation (\ref{SUSYorthogonalityLandx}), we obtain
$L_a^2+C_{\alpha\beta}L_{\alpha}L_{\beta}=\Lambda_a^2+C_{\alpha\beta}\Lambda_{\alpha}\Lambda_{\beta}+{(I/2)^2}$,
and then the Casimir index for $L_a$, $L_{\alpha}$  is given by
$J=n+I/2$
 ($n$  corresponds to the Landau level index).
Therefore, the energy eigenvalue  is derived as
\begin{gather*}
E_n=
\frac{1}{2MR^2}\left(n\left(n+I+\frac{1}{2}\right)+\frac{I}{4}\right).
\end{gather*}
The degeneracy in $n$-th Landau level is\footnote{In the original
system, the energy eigenvalue of  $n$-th Landau level is
 $E_n=\frac{1}{2M}(n(n+I+{1})+\frac{I}{2})$ and the degeneracy is
$D_n=2n+I+1$. The degeneracy in the SUSY system is almost doubly
degenerate compared to the original ``bosonic'' system due to the
existence of the ``fermionic'' counterpart. Especially, at
$I\rightarrow \infty$ in the LLL,
 $D_{n=0}\rightarrow N_{\Phi}=I$ in the original model, while $D_{n=0}\rightarrow 2N_{\Phi}=2I$ in the SUSY model.}
\begin{gather*}
D_n=4n+2I+1.
\end{gather*}
In the lowest Landau level (LLL) $n=0$, the energy is
\begin{gather}
E_{LLL}=\frac{I}{8MR^2}=\frac{B}{4M}, \label{LLLSUSYenergy}
\end{gather}
 and there are  $2I+1$ degenerate eigenstates that consist of
\begin{gather}
u_{m_1,m_2}=\sqrt{\frac{I!}{m_1!m_2!}}u^{m_1}v^{m_2},\qquad
\eta_{n_1,n_2}=\sqrt{\frac{I!}{n_1!n_2!}}u^{n_1}v^{n_2}\eta,\label{susymonohar}
\end{gather}
with the constraints $m_1+m_2=I$ and $n_1+n_2=I-1$. $u_{m_1,m_2}$
is a Grassmann even quantity, while $v_{n_1,n_2}$ is a Grassmann
odd quantity. The eigenvalues of $L_3$ for $u_{m_1,m_2}$ and
$\eta_{n_1,n_2}$ are explicitly given by $I,I-1,\dots,-I+1,-I$ and
$I-1/2,I-3/2,\dots,-I+1/2$, respectively, and thus  dif\/fer by
$1/2$. Since  $u_{m_1,m_2}$ and $v_{n_1,n_2}$ are
 related by the transformation generated by the fermionic operators $L_{\alpha}$, they  are
regarded as SUSY partners and  named the supermonopole harmonics;
$\eta_{n_1,n_2}$  are the   ``fermionic'' counterpart  of the
original ``bosonic'' monopole harmonics~$u_{m_1,m_2}$. The
orthonormal relations are
\begin{gather}
\int_{S^{2|2}} d\Omega_{2|2} u_{m_1,m_2}^* u_{m'_1,m'_2}=\frac{4\pi I}{I+1}\delta_{m_1,m'_1}\delta_{m_2,m'_2},\nonumber\\
\int_{S^{2|2}} d\Omega_{2|2} \eta_{n_1,n_2}^* \eta_{n'_1,n'_2}= 4\pi \delta_{n_1,n'_1}\delta_{n_2,n'_2},\nonumber\\
\int_{S^{2|2}} d\Omega_{2|2} u_{m_1,m_2}^*
\eta_{n_1,n_2}=0,\label{normalizationonsupersphere}
\end{gather}
with $d\Omega_{2|2}=d\omega_2d\theta_1d\theta_2$; $d\omega_2$ is
the area element of  two-sphere.

It should be also noted that  the $u_{m_1,m_2}$ and $v_{n_1,n_2}$
are constructed by products of the components of  SUSY Hopf
spinor, so  the $OSp(1|2)$ generators are ef\/fectively
represented  as
\begin{gather}
L_a=\psi^t\tilde{l}_{\alpha}\frac{\partial}{\partial\psi},\qquad
L_{\alpha}= \psi^t\tilde{l}_{\alpha}\frac{\partial}{\partial\psi},
\label{LLLorbitalangu}
\end{gather}
where $\tilde{l}_a$ and $\tilde{l}_{\alpha}$ are def\/ined by
equation~(\ref{complexrepospgene}). The supermonopole charge is
measured by the operator:
\begin{gather}
\hat{I}=u\frac{\partial}{\partial u}+{v}\frac{\partial}{\partial
v}+\eta\frac{\partial} {\partial \eta}. \label{operatormonochar}
\end{gather}
Complex variables never appear in the LLL bases
(\ref{susymonohar}), and they are replaced by  derivatives:
\begin{gather}
\psi^{*}=(u^*,v^*,\eta^*)^t \rightarrow
\frac{1}{I}\frac{\partial}{\partial\psi}=\frac{1}{I}\left(\frac{\partial}{\partial
u}, \frac{\partial}{\partial v},\frac{\partial}{\partial
\eta}\right)^t, \label{effectivcepsiinLLL}
\end{gather}
as suggested by  equations (\ref{normalizationonsupersphere}).
This substitution implies that $\psi$ and $\psi^*$ no longer
commute  each other;   this gives rise to NCG in LLL.

Here, we make some comments  about peculiar properties of the
present supersymmetry.
The Hamiltonian (\ref{SUSYoneHam}), which is equal to the $OSp(1|2)$ Casimir operator up to constant,  appa\-rent\-ly commutes with the supercharges $L_{\alpha}$
, and, in this sense,     the present model possesses a
supersymmetry.
However, there are some dif\/ferences between the present SUSY
model and
 conventional SUSY quantum mechanics.
First of all, the present model is def\/ined on  supermani\-fold,
while  basemanifolds for   SUSY quantum mechanics are usually
taken to be bosonic. Then, in this model, energy eigenfunctions
generally depend on  Grassmann odd coordinates as well as
Grassmann even coordinates. (In this sense, our wavefunctions are
something like superf\/ields.) Second, the supercharges are not
nilpotent: $L_{\theta_1}^2 = (L_x+iL_y)/4 $ and
$L_{\theta_2}^2=-(L_x-iL_y)/4$   as suggested by the $OSp(1|2)$
algebra (\ref{ospalgebrarelation}).  Thus, the square of the
supercharges acts as  the ladder operators for  $SU(2)$ spin 1,
and the supercharges themselves are regarded as   ladder operators
for $SU(2)$ spin 1/2. Last, the Hamiltonian (\ref{SUSYoneHam}) is
not given by the anticommutator of  supercharges, so the lowest
energy (LLL energy) is not  zero but f\/inite
(\ref{LLLSUSYenergy}). Similarly,   bosonic degrees of freedom do not exactly equal to fermionic ones but dif\/fer~by 1.

\subsection[The spherical SUSY Laughlin wavefunction and excitations {[35]}]{The spherical SUSY Laughlin wavefunction and excitations \cite{PRL94206802}} \label{SphSUSYLaugh}

 Now, we are ready to  discuss  many-body problem.  First, we construct the groundstate wavefunction of the SUSY QHE.
The original Laughlin wavefunction on  a two-sphere was  given by
\begin{gather}
\Phi= \prod_{i<j}^N (\phi_i^t C\phi_j)^m    =
\prod_{i<j}^N(u_i v_j-v_i u_j)^m, \label{originalsphereLaugh}
\end{gather}
where $\phi=(u,v)^t$ represents the original $SU(2)$ Hopf spinor
and $N$ represents the total number of  particles and $m$ denotes
integer \cite{HaldanePRL51}. Thus, $\Phi$ is constructed by the
product of  $SU(2)$ singlets of two Hopf spinors. Then, it would
be natural to   adopt product of $OSp(1|2)$ singlets made of two
super Hopf spinors as a SUSY Laughlin wavefunction. However, as
discussed above,  we cannot use complex variables to construct
$OSp(1|2)$ singlet in  LLL. Fortunately,  the complex and the
original representations are unitary equivalent, and hence  it is
possible  to construct a singlet  of two super Hopf spinors
without introducing  complex variables. Thus, the spherical SUSY
Laughlin wavefunction is constructed as
\begin{gather}
\Psi= \prod_{i<j}^N (\psi_i^t \mathcal{R}\psi_j)^m    =
\prod_{i<j}^N(u_iv_j-u_jv_i-\eta_i\eta_j)^m.
\label{sphericalSUSYLauhglinwave}
\end{gather}
Since the SUSY Laughlin wavefunction is  invariant under the
$OSp(1|2)$ SUSY transformation, the super partner of the SUSY
Laughlin wavefunction does not exist. Acting the monopole charge
operator (\ref{operatormonochar}) to the SUSY wavefunction, one
may see that the monopole charge $I$ is related to $N$ and $m$ as
\begin{gather}
I=m(N-1). \label{relationNI}
\end{gather}
There may be  two choices to def\/ine the f\/illing factor $\nu$
in the SUSY QHE: $N/N_{\Phi}$ or $N/D$. In the thermodynamic limit
($N,I,R \rightarrow \infty$ with the magnetic length $\ell_B$
f\/ixed), these two def\/initions are dif\/ferent; $N/D
\rightarrow N/2I=N/2N_{\Phi}$, unlike the original
QHE{\footnote{In the original QHE,  these two def\/initions
coincide in the thermodynamic limit: $N/D \rightarrow
N/I=N/N_{\Phi}$.}}. It is convenient to  use the def\/inition
\begin{gather*}
\nu=\frac{N}{N_{\Phi}}=\frac{N}{I}.
\end{gather*}
The f\/illing factor for the SUSY Laughlin wavefunction reads as
$N/m(N-1)$, which tends to $\nu=1/m$ in the thermodynamic limit.

It is also possible to construct a pseudo-potential Hamiltonian
whose zero-energy eigenstate  is the SUSY Laughlin wavefunction.
The SUSY Laughlin wavefunction is $OSp(1|2)$ symmetric, and   does
not have any components whose 3rd component of the two-body
angular momentum eigenvalue is greater than $m(N-2)=I-m$. Then,
the  SUSY Laughlin wavefunction does not contain any components
whose two-body $OSp(1|2)$ Casimir index $J$ is greater than $I-m$.
From this observation, the  pseudo-potential Hamiltonian  is
derived as
\begin{gather}
\hat{V}=\sum_{J=I-m+1/2,I-m+1,\dots,I}V_J\cdot
P_J(L_a(i)L_a(j)+C_{\alpha\beta}L_{\alpha}(i)L_{\beta}(j)),
\label{hardcoreHam}
\end{gather}
where coef\/f\/icients  $V_J$ are taken to be positive, and $P_J$
is   given by
\begin{gather*}
P_J(L_a(i)L_a(j)+C_{\alpha\beta}L_{\alpha}(i)L_{\beta}(j))\nonumber\\
\qquad{} = \prod_{J'\neq J}
\frac{(L_a(i)+L_a(j))^2+C_{\alpha\beta}( L_{\alpha}(i)+L_{\alpha}(j) )( L_{\beta}(i)+L_{\beta}(j))-J'(J'+\frac{1}{2}) }{J(J+\frac{1}{2}) -J'(J'+\frac{1}{2})}\nonumber\\
\qquad{}=  \prod_{J'\neq J} \frac{2L_a(i)L_a(j)+2C_{\alpha\beta}
L_{\alpha}(i)L_{\beta}(j)+ \frac{I}{2}(I+1)-J'(J'+\frac{1}{2})
}{J(J+\frac{1}{2})-J'(J'+\frac{1}{2})}.
\end{gather*}
$P_J$ denotes the projection operator  to the subspace of
two-body $OSp(1|2)$ Casimir index $J$. With positive
coef\/f\/icients $V_J$, the lowest energy eigenvalue of the
Hamiltonian (\ref{hardcoreHam}) is zero, and  the pseudo-potential
Hamiltonian does not have any component $J< I-m$. Therefore,  the
SUSY Laughlin wavefunction is the zero-energy exact groundstate of
the Hamiltonian\footnote{It is reported that in  scalar f\/ield
theories on  supersphere  ef\/fective potentials are not generally
bounded below, so the groundstates are not stable
\cite{hep-th/0409257}. However,  such  problem cannot be applied
to SUSY QHE, since the pseudo-potential Hamiltonian \eqref{hardcoreHam} has the lowest
eigenvalue and the energies are bounded.}.

Quasi-hole (= vortex) and quasi-particle (= anti-vortex) operators
are respectively constructed as
\begin{gather*}
A(\chi)^{\ddagger}=\prod_i \psi_i \mathcal{R}\chi =\prod_i (b v_i-a u_i-\xi\eta_i ),\\
A(\chi)=\prod_i \chi^{\ddagger}\mathcal{R}^t
\frac{\partial}{\partial \psi_i}=\prod_i
\left(b^*\frac{\partial}{\partial v_i}-a^*\frac{\partial}{\partial
u_i} -\xi^* \frac{\partial}{\partial\eta_i}\right),
\end{gather*}
where $\chi\equiv (a,b,\xi)^t$ is a  normalized constant spinor,
which specif\/ies the position at which the quasi-hole
(quasi-particle) is created on the supersphere by  relations:
$\Omega_a=2\chi^{\ddagger}l_a\chi$ and
$\Omega_{\alpha}=2\chi^{\ddagger}l_{\alpha}\chi$. Their
commutation relations are
\begin{gather*}
[A(\chi),A(\chi)^{\ddagger}]=1,\qquad
[A(\chi),A(\chi')]=[A^{\ddagger}(\chi),A^{\ddagger}(\chi')]=0.
\end{gather*}
Similarly, the commutation relation between the quasi-hole
operator  and  the $OSp(1|2)$ operators in the direction
$(\Omega_a,\Omega_{\alpha})$ is given by
\begin{gather*}
[\Omega_a(\chi)L_a+C_{\alpha\beta}\Omega_{\alpha}(\chi)L_{\beta},
A({\chi})]=\frac{N}{2}A(\chi).
\end{gather*}
This implies that the creation of  quasi-hole increases  the
angular momentum in the direction of the point
$(\Omega_a,\Omega_{\alpha})$ by $N/2$. Physically, it is
understood as follows. The creation of quasi-hole pushes the
particles on the SUSY Laughlin state downward from  the point
$(\Omega_a,\Omega_{\alpha})$, so the charge def\/icit which we
identify quasi-particle is generated  at the point. The relation
(\ref{relationNI}) suggests that the excess of   unit magnetic
f\/lux, $\delta I=1$, induces  excitation with fractional charge
$e^*={1}/{m}$. Thus, the   quasi-particle excitation in SUSY QHE
at $\nu=1/m$ carries the fractional charge $1/m$ as in the
original QHE  \cite{HaldanePRL51}, and  the fractional charge is
induced  by bosonic and fermionic  Hall currents as suggested by
equations~(\ref{SUSYHalllawonsphere}).

\section[Emergence of  non-anti-commutative geometry {[34,35]}]{Emergence of  non-anti-commutative geometry \cite{NPB70994,PRL94206802}}

Originally, $x_a$ and $\theta_{\alpha}$ were the classical
coordinates   on  supersphere and  not operators, while, in the
LLL, they  are ef\/fectively regarded as operators. It is because,
in the LLL limit 
($M\rightarrow 0$)  the covariant angular momenta can be neglected
(see equations~(\ref{covariantangmomentaxtheta})), and $x_a$ and
$\theta_{\alpha}$   are  reduced to the $OSp(1|2)$ operators  as
indicated  by equation~(\ref{llambdax}):
\begin{gather}
(x_a,\theta_{\alpha})\rightarrow (X_a,\Theta_{\alpha})\equiv
-{\alpha}(L_a,L_{\alpha}), \label{LLLreductionxtoL}
\end{gather}
with $\alpha=2R/I$. Thus, in the LLL,  $x_a$ and $\theta_{\alpha}$
become
 operators that satisfy  the  SUSY NC algebra:
\begin{gather}
[X_a,X_b]=-i\alpha\epsilon_{abc}X_c, \qquad
[X_a,\Theta_{\alpha}]=-\frac{\alpha}{2}(\sigma_a)_{\beta\alpha}\Theta_{\beta},
\nonumber\\
\{\Theta_{\alpha},\Theta_{\beta}\}=-\frac{\alpha}{2}(C\sigma_a)_{\alpha\beta}X_a.
\label{fuzzysuperspherealgebra}
\end{gather}
The f\/irst relation manifests the  noncommutativity in the LLL,
and the second relation suggests the non-trivial ``coupling''
between the bosonic and fermionic operators. The latter relation
ref\/lects   the non-anti-commutative geometry in the SUSY LLL.
The fuzzy super manifold introduced by the algebraic relation
(\ref{fuzzysuperspherealgebra}) is known as  the fuzzy supersphere
\cite{hep-th/9507074,math-ph/9804013}\footnote{The fuzzy
supersphere is a classical solution of supermatrix model
\cite{hep-th/0311005}. A nice review of  mathematics and physical
applications  of
 fuzzy sphere and fuzzy supersphere is found in~\cite{hep-th/0511114}.}.
Thus,  SUSY NCG  are nicely realized in the formulation of the
SUSY QHE.
Alternatively, one may  f\/ind  the emergence of fuzzy supersphere
by the following
 derivation.
Complex variables are regarded as   derivatives in the LLL
(\ref{effectivcepsiinLLL}), and the Hopf map (\ref{superHopfmap2})
is reduced to
\begin{gather}
X_a=-\alpha\psi^t \tilde{l_a}\frac{\partial}{\partial\psi},\qquad
\Theta_{\alpha}=-\alpha\psi^t
\tilde{l_{\alpha}}\frac{\partial}{\partial\psi}.
\label{LLLHopfmap}
\end{gather}
Apparently, $X_a$ and $\Theta_{\alpha}$ satisfy the algebra  of
fuzzy supersphere. By comparison of  equation~(\ref{LLLHopfmap})
and equation~(\ref{LLLorbitalangu}),  the equivalence between
$(L_a,L_{\alpha})$ and $(X_a,\Theta_{\alpha})$ in LLL
(\ref{LLLreductionxtoL})  is also conf\/irmed.
Equations~(\ref{fuzzysuperspherealgebra}) imply that  the super
Hall currents $I_a=\frac{d}{dt}X_a$,
$I_{\alpha}=\frac{d}{dt}\Theta_{\alpha}$  satisfy the relations,
\begin{gather}
I_a=-i[X_a,V]=(\alpha R)^2\epsilon_{abc}B_b E_c- i\tfrac{1 }{2}
(\alpha R)^2 (\sigma_a C)_{\alpha\beta}
B_{\alpha}E_{\beta},\nonumber\\
I_{\alpha}=-i[\Theta_{\alpha},V]=i\tfrac{1}{2} (\alpha R)^2
(\sigma_a)_{\beta\alpha} B_a E_{\beta} +i\tfrac{1}{2} (\alpha R)^2
(\sigma_a)_{\beta\alpha} B_{\beta}
E_a,\label{SUSYHalllawonsphere}
\end{gather}
where $E_a=-\partial_a V$,
$E_{\alpha}=C_{\alpha\beta}\partial_{\beta} V$. With
equations~(\ref{SUSYHalllawonsphere}), it is checked that  the
super Hall currents are orthogonal to the super electric f\/ields
and  the super magnetic f\/ields, respectively:
\begin{gather}
E_a I_a +C_{\alpha\beta}E_{\alpha} I_{\beta}=0,\qquad 
B_a I_a +C_{\alpha\beta} B_{\alpha}
I_{\beta}=0.\label{totalorthogEand}
\end{gather}
Around the north-pole of the fuzzy supersphere $X_3 \approx \alpha
I/2$,  the SUSY NC algebra (\ref{fuzzysuperspherealgebra}) is
reduced to  that on the  NC superplane,
\begin{gather}
[X_i,X_j]=i\epsilon_{ij}\ell_B^2,\qquad
[X_i,\Theta_{\alpha}]=0,\qquad
\{\Theta_{\alpha},\Theta_{\beta}\}=(\sigma_1)_{\alpha\beta}\ell_B^2.\label{planarNC}
\end{gather}
Such planar reductions were well examined by the
In\"on\"u--Wigner contraction technique in more general contexts
\cite{hep-th/0306251}.
 As we shall see below,
the planar SUSY QHE naturally manifests the  planar SUSY NC
algebra in LLL.

\section[The planar SUSY quantum Hall effect]{The planar SUSY quantum Hall ef\/fect}\label{planarsystem}

\subsection[The generators on the superplane and stereographic projection {[40,41]}]{The generators on the superplane and stereographic projection \cite{PRD72105017,0705.4527}}

Using the In\"on\"u--Wigner contraction, we derive the symmetry
generators  on the superplane from the $OSp(1|2)$ generators. We
apply a symmetric scaling to the $OSp(1|2)$ generators as
\begin{gather*}
L_i \rightarrow \epsilon T_i,\qquad L_{\alpha} \rightarrow
\epsilon T_{\alpha},\qquad L_3\rightarrow L_{\perp}.
\end{gather*}
By taking the limit $\epsilon\rightarrow 0$, the $OSp(1|2)$ SUSY
commutation relations are reduced to the translation and rotation
algebras on the superplane,
\begin{gather}
[T_i,T_j]=0,\qquad [T_i,L_{\perp}]=-i\epsilon_{ij}T_j,\nonumber\\
[T_i,T_{\alpha}]=0,\label{superplanealge}\\
\{T_{\alpha},T_{\beta}\}=0,\qquad [T_{\alpha},L_{\perp}]=\pm
\tfrac{1}{2}T_{\alpha},\nonumber
\end{gather}
where, in the latter equation,  $+$ and $-$ correspond to
$\alpha=1$ and $\alpha=2$, respectively.
 The dif\/ferential operators that satisfy (\ref{superplanealge}) denote  the
  translation generators  and the perpendicular angular momentum  on the superplane. They
 are explicitly represented as
\begin{gather*}
T_i=-i\partial_i,\qquad T_{\alpha}=-i\partial_{\alpha},\qquad
L_{\perp}= z\frac{\partial}{\partial
z}-z^*\frac{\partial}{\partial
z^*}+\tfrac{1}{2}\theta\frac{\partial}{\partial\theta}-\tfrac{1}{2}\theta^*\frac{\partial}{\partial\theta^*}.
\end{gather*}
The f\/irst two terms in $L_{\perp}$ denote the conventional
orbital angular momentum and count the dif\/ference between the
powers of  $z$ and $z^*$. Essentially, $z$ corresponds to  the
right-handed orbital rotation, and $z^*$ the left-handed orbital
rotation. Similarly,  $\theta$ may be  regarded as the
right-handed spin rotation and $\theta^*$ the left-handed spin
rotation. Indeed, the factor $1/2$ in front of the  last two terms
in $L_{\perp}$ implies $\theta$ and $\theta^*$ carry the spin-up
and  spin-down degree of freedom, respectively.

Introducing the super stereographic coordinates $(z, \theta)$
\begin{gather}
z\equiv \frac{v}{u}=
\frac{x_1+ix_2}{R+x_3}\biggl(1+\frac{1}{2R(R+x_3)}\theta
C\theta\biggr),\qquad \theta\equiv \frac{\eta}{u}=\theta_1+z
\theta_2, \label{defstereog}
\end{gather}
 we simply express  the SUSY Hopf spinor  as
\begin{gather*}
\psi=\frac{1}{\sqrt{1+zz^*+\theta\theta^*}}
\begin{pmatrix}
1 \\
z\\
\theta
\end{pmatrix}.
\end{gather*}
The supermonopole harmonics are also rewritten as
\begin{gather*}
u_{m_1,m_2}= \sqrt{\frac{I!}{{m_1 ! m_2 !}}} z^{m_2}
\biggl(\frac{1}{1+zz^*+\theta\theta^*}\biggr)^{\frac{I}{2}},\qquad
\eta_{n_1,n_2}=\sqrt{\frac{I!}{{n_1! n_2 !}}} z^{n_2}\theta
\biggl(\frac{1}{1+zz^*+\theta\theta^*}\biggr)^{\frac{I}{2}},
\end{gather*}
and, in  the  thermodynamic limit,   they  become
\begin{gather}
\phi_{m}= \sqrt{\frac{2^{m+1}}{\pi m!}} z^{m}
e^{-zz^*-\theta\theta^*},\qquad
\psi_{m-\frac{1}{2}}=\sqrt{\frac{2^{m}}{\pi (m-1)!}} z^{m-1}\theta
e^{-zz^*-\theta\theta^*}.\label{LLLbasecoordinate}
\end{gather}
Their  coef\/f\/icients  are chosen to satisfy the orthonormal
conditions:
\begin{gather*}
 \int dz dz^* d\theta d\theta^*  \phi_m^{*}\phi_m' = \int dz dz^* d\theta d\theta^*  \psi_{m-1/2}^{*}\psi_{m'-1/2}=\delta_{m m'},\\
 \int dz dz^* d\theta d\theta^*  \phi_m^{*}\psi_{m'-1/2}=0.
\end{gather*}
As found in equations~(\ref{LLLbasecoordinate}),   the complex
variables $z^*$ and $\theta^*$ do not appear in the LLL up to the
exponential. In the LLL,
 $z^*$ and $\theta^*$  are equivalent to  derivatives:
\begin{gather}
(z^*, \theta^*)\rightarrow \left(-\frac{\partial}{\partial
z},-\frac{\partial}{\partial\theta}\right).
\label{effectivez*theta*}
\end{gather}
It is apparent that  operations of  the derivatives   to
(\ref{LLLbasecoordinate}) are same  as of  the complex variables.

\subsection[The planar SUSY Landau problem {[40]}]{The planar SUSY Landau problem \cite{PRD72105017}}\label{subsecplanarlandaupro}

As the planar SUSY Hamiltonian  we adopt the following operator
\begin{gather}
H=-\frac{1}{2M}(D_i^2+C_{\alpha\beta}D_{\alpha}D_{\beta}),
\label{planarSUSYHam}
\end{gather}
where $D_i$ and $D_{\alpha}$ denote the SUSY covariant derivatives
def\/ined
 by $D_i=\partial_i-iA_i$ and $D_{\alpha}=\partial_{\alpha}-iA_{\alpha}$, with
$A_i={B}/{2}\epsilon_{ij}x_j$ and
$A_{\alpha}={B}/{2}(\sigma_1)_{\alpha\beta}\theta_{\beta}$.
Identifying the stereographic coordina\-tes~(\ref{defstereog})
with    $x_i$ and $\theta_{\alpha}$:
\begin{gather}
 z= \frac{1}{2\ell_B}(x+iy) ,\qquad z^*= \frac{1}{2\ell_B}(x-iy) ,\qquad
 \theta= \frac{1}{\sqrt{2}\ell_B}\theta_1 ,\qquad \theta^*= \frac{1}{\sqrt{2}\ell_B}\theta_2,\label{defzandthetabyxy}
\end{gather}
it is  easily shown  that  LLL bases (\ref{LLLbasecoordinate}) are
zero-energy degenerate groundstates of the SUSY Hamiltonian.

\begin{figure}[t]
  \centerline{\includegraphics[height=43mm]{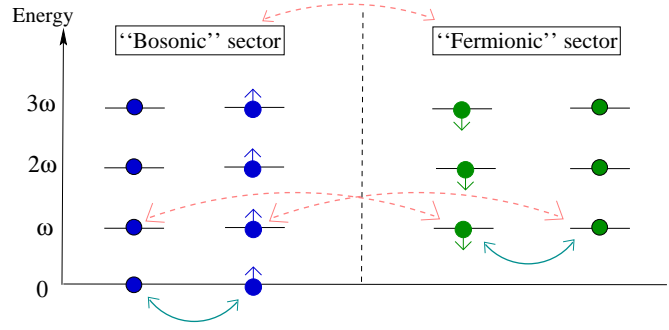}}

  \caption{The ``balls'' correspond to the
states given by equation~(\ref{SUSYmultiplane}). The  solid curved
arrows represent the  $Q$-SUSY transformations, while the  dotted
curved arrows represent the
 $\tilde{Q}$-SUSY transformations.}
  \label{planarSUSYN=2.fig}
\end{figure}

The SUSY covariant derivatives $D_i$ and $D_{\alpha}$ satisfy the
algebra,
\begin{gather}
[D_i,D_j]=iB\epsilon_{ij},\qquad
\{D_\alpha,D_{\beta}\}=-B{(\sigma_1)}_{\alpha\beta},\qquad
[D_i,D_{\alpha}]=0. \label{SUSYrelD}
\end{gather}
The SUSY center-of-mass coordinates are constructed as
\begin{gather}
 X_i=x_i+i\ell_B^2 D_j,\qquad \Theta_{\alpha}=\theta_{\alpha}+\ell_B^2 D_{\alpha},
\end{gather}
 and    satisfy the SUSY NC relations,
\begin{gather}
[X_i,X_j]=i\ell_B^2\epsilon_{ij}, \qquad
\{\Theta_{\alpha},\Theta_{\beta}\}=\ell_B^2(\sigma_1)_{\alpha\beta},\qquad
[X_i,\Theta_{\alpha}]=0. \label{SUSYNConplane}
\end{gather}
In the LLL limit\footnote{The LLL limit is formally realized by
neglecting the SUSY covariant derivatives $D_i$ and $D_{\alpha}$.}
($M\rightarrow 0$), $x_i$ and $\theta_{\alpha}$ are reduced to
$X_i$ and $\Theta_{\alpha}$ respectively, so the SUSY NC relations
(\ref{SUSYNConplane}) are realized in the planar SUSY  QHE as
expected. Equations~(\ref{defzandthetabyxy}) and
equations~(\ref{SUSYNConplane}) suggest that $z$ and $z^*$  are no
longer commutative  but noncommutative, and  similarly $\theta$
and $\theta^*$ are no longer anti-commutative but
non-anti-commutative
 in the LLL.  This observation  is consistent with the substitution (\ref{effectivez*theta*}). From  two-sets of SUSY commutation relations (\ref{SUSYrelD})--(\ref{SUSYNConplane}),  two sets of  bosonic and fermionic raising and lowering operators are naturally def\/ined:
\begin{gather*}
 a= -i\frac{\ell_B}{\sqrt{2}}(D_x+iD_y), \qquad a^{\dagger}=-i\frac{\ell_B}{\sqrt{2}}(D_x-iD_y),\qquad \alpha=i\ell_B D_{\theta_2},\qquad \alpha^{\dagger}=i\ell_B D_{\theta_1},
\end{gather*}
 and
\begin{gather*}
b=\frac{1}{\sqrt{2}\ell_B}(X-iY),\qquad
b^{\dagger}=\frac{1}{\sqrt{2}\ell_B}(X+iY),\qquad
\beta=\frac{1}{\ell_B}\Theta_2,\qquad
\beta^{\dagger}=\frac{1}{\ell_B}\Theta_1.
\end{gather*}
 With such SUSY raising and lowering operators, one may construct two kinds of supercharges:
\begin{gather*}
(Q, Q^{\dagger})=(a^{\dagger}\alpha,  \alpha^{\dagger}a),\qquad
(\tilde{Q},\tilde{Q}^{\dagger})=(b^{\dagger}\beta,
\beta^{\dagger}b).
\end{gather*}
The f\/irst set  and the second set  are anti-commutative  each
other,  and  these sets of supercharges   generate two independent
SUSY transformations that we call $Q$-SUSY and $\tilde{Q}$-SUSY.
With use of  $Q$-SUSY generators, the Hamiltonian
(\ref{planarSUSYHam}) is rewritten as
\begin{gather*}
 H={\omega} \{Q,Q^{\dagger}\}={\omega}(a^{\dagger}a+\alpha^{\dagger}\alpha).
\end{gather*}
Its eigenenergy is
\begin{gather*}
E_n=\omega n,
\end{gather*}
where $n$ takes integer that specif\/ies the SUSY Landau level.
Since the Hamiltonian  commutes with  $\tilde{Q}$ and
$\tilde{Q}^{\dagger}$ in addition to $Q$ and $Q^{\dagger}$,  the
planar SUSY model possesses $\mathcal{N}=2$ SUSY in total. The
$\mathcal{N}=2$ SUSY multiplets are constructed by acting the
operators
\begin{gather}
\frac{1}{\sqrt{n!m!}} {a^{\dagger}}^n{b^{\dagger}}^m,\qquad
\frac{1}{\sqrt{n!(m-1)!}} {a^{\dagger}}^n \beta^{\dagger}{b^{\dagger}}^{m-1},\qquad \frac{1}{\sqrt{(n-1)!m!}} \alpha^{\dagger}{a^{\dagger}}^{n-1}{b^{\dagger}}^m,\nonumber\\
\frac{1}{\sqrt{(n-1)!(m-1)!}}
\alpha^{\dagger}{a^{\dagger}}^{n-1}\beta^{\dagger}{b^{\dagger}}^{m-1},
\label{SUSYmultiplane}
\end{gather}
to the vacuum (see  Fig.~\ref{planarSUSYN=2.fig}). Specif\/ically,
the LLL $(n=0)$-sector  consists of
\begin{gather}
|\phi_m\rangle =\frac{1}{\sqrt{n!m!}}
{a^{\dagger}}^n{b^{\dagger}}^m|0\rangle ,\qquad
|\psi_{m-\frac{1}{2}}\rangle =\frac{1}{\sqrt{n!(m-1)!}}
{a^{\dagger}}^n \beta^{\dagger}{b^{\dagger}}^{m-1}|0\rangle.
\label{LLLbasesket}
\end{gather}
These are supermultiplet related by $\tilde{Q}$-SUSY. Thus, while
the LLL is the ``vacuum'' of the $Q$-SUSY, there still exist
$\mathcal{N}=1$ SUSY degeneracies due to   $\tilde{Q}$-SUSY. The
LLL wavefunctions (\ref{LLLbasecoordinate}) are reproduced from
equations~(\ref{LLLbasesket}) with the  vacuum
$\phi_0=\frac{1}{\sqrt{\pi}}e^{-zz^*-\theta\theta^*}$.

The perpendicular angular momentum $L_{\perp}$ is expressed by the
SUSY creation and annihilation operators as
\begin{gather*}
L_{\perp}=\big(b^{\dagger}b+\tfrac{1}{2}\beta^{\dagger}\beta\big)
-\big(a^{\dagger}a+\tfrac{1}{2}\alpha^{\dagger}\alpha\big),
\end{gather*}
and the commutation relations between the supercharges and
$L_{\perp}$ are
\begin{gather*}
[L_{\perp}, Q]=-\tfrac{1}{2}Q,\qquad
[L_{\perp},Q^{\dagger}]=\tfrac{1}{2}Q^{\dagger},\qquad
[L_{\perp},\tilde{Q}]=\tfrac{1}{2}\tilde{Q},\qquad
[L_{\perp},\tilde{Q}^{\dagger}]=-\tfrac{1}{2}\tilde{Q}^{\dagger}.
\end{gather*}
Thus, the supercharges are spin 1/2 operators. The magnitudes of
the perpendicular angular momenta for the $\mathcal{N}=2$ SUSY
multiplets (\ref{SUSYmultiplane}) are respectively given by
$(m-n)$, $(m-n-\frac{1}{2})$, $(m-n+\frac{1}{2})$ and $(m-n)$. The
multiplets possess  the same orbital angular momentum: $(m-n)$,
while their spins are dif\/ferent: $0$, $1/2$, $0$ and $-1/2$
(Fig.~\ref{planarSUSYN=2.fig}).

\subsection[The SUSY Laughlin wavefunction {[41]}]{The SUSY Laughlin wavefunction \cite{0705.4527}}\label{susylaughplane}

\begin{figure}[t]

\centerline{\includegraphics[width=15.5cm]{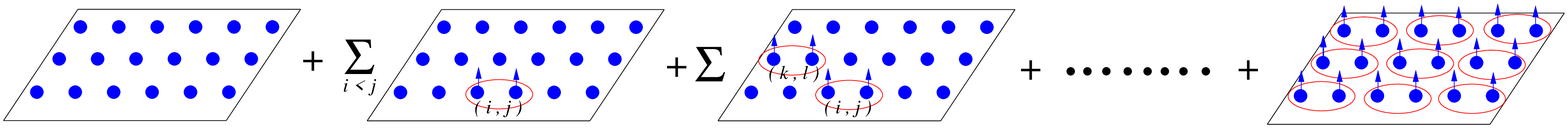}}

  \caption{The graphical representation for the expansion (\ref{expansionSUSYLlin}).
Each  circle represents the $p$-wave pairing  on the Laughlin
state. In the $n$-th component wavefunction of the expansion,
 the  pairing operator acts to the Laughlin state $n - 1$ times and constitute
 $n - 1$ spin-polarized  $p$-wave pairings. \label{expansionSUSYLlin.fig}}
  \end{figure}

In the thermodynamic limit, the spherical SUSY Laughlin
wavefunction (\ref{sphericalSUSYLauhglinwave}) is transformed to
the planar  SUSY Laughlin wavefunction:
\begin{gather*}
\Psi=\prod_{i<j}^N (z_i-z_j+\theta_i\theta_j)^m e^{-\sum_i
(z_iz^*_i+\theta_i\theta_i^*)},
\end{gather*}
where $z$ and $\theta$ are  the stereographic coordinates. To
explore its physical meaning, it is important to notice that  the
SUSY Laughlin wavefunction is rewritten in the form:
\begin{gather}
\Psi=\exp\left(m \sum_{i<j} \frac{\theta_i\theta_j}{z_i-z_j}\right)\cdot \Phi=\Phi+m\sum_{i<j}\frac{\theta_i\theta_j}{z_i-z_j}\Phi \nonumber\\
\phantom{\Psi=}{}
+\frac{m^2}{2}\left(\sum_{i<j}\frac{\theta_i\theta_j}{z_i-z_j}\right)^2\Phi
+\cdots+\frac{m^{\frac{N}{2}}}{(N/2)!}\theta_1\theta_2\cdots
\theta_N\!\cdot \!
Pf\left(\frac{1}{z_i-z_j}\right)\Phi,\label{expansionSUSYLlin}
\end{gather}
where  $\Phi$ denotes  the planar version of the original Laughlin
wavefunction (\ref{originalsphereLaugh}),
\begin{gather*}
\Phi=\prod_{i<j}(z_i-z_j)^m e^{-\sum_i
(z_iz_i^*+\theta_i\theta_i^*)}.
\end{gather*}
In the second equation of equation~(\ref{expansionSUSYLlin}), we
expanded the exponential in terms of Grassmann quantity,
$\sum\limits_{i<j}{\theta_i\theta_j}/({z_i-z_j})$, which we call
the pairing operator hereafter.
 Since $\theta$ carries spin $1/2$ degree of freedom, the numerator $\theta_i\theta_j$ acts to attach spin $1/2$ to each of  the original Laughlin spinless particles $i$ and $j$. Meanwhile, the denominator
$1/(z_i-z_j)$ is a solution of the 2D Schr\"odinger equation with
attractive contact  interaction, and represents a $p$-wave pairing
state of  $i$ and $j$ particles.  Then, in total, the pairing
operator ${\theta_i\theta_j}/({z_i-z_j})$ may be regarded as an
operator that forms
  a spin-polarized  $p$-wave  pairing state of  $i$, $j$ particles on  the Laughlin state.
With this interpretation, the expansion~(\ref{expansionSUSYLlin})
now has the following physical meaning. Apparently, the 1st
component of the expansion  is the original Laughlin wavefunction.
In the 2nd component, the pairing operator acts to the original
Laughlin wavefunction once, and  one $p$-wave pairing state is
generated on the Laughlin state. Similarly, in the 3rd component,
the pairing operator acts on the Laughlin function twice, and  two
$p$-wave pairings  are generated on the Laughlin state. Repeating
this procedure, we f\/inally arrive at the state where all
particles form $p$-wave pairings  with polarized spins
(Fig.~\ref{expansionSUSYLlin.fig}). This  state is nothing but
Moore--Read state~\cite{NPB360362} that  was proposed as  a
candidate  groundstate at even denominator f\/illings
\cite{PRL663205}\footnote{Especially,  the Moore--Read state is a
most promising candidate for  the QH groundstate at the f\/illing
$5/2$, where  $p$-wave pairing bosons condense to form a
``bosonic'' QH liquid.}.  Indeed,  Pfaf\/f\/ian form proposed by
Moore and Read    appears  as  the last component wavefunction  in
the expansion~(\ref{expansionSUSYLlin}). Thus, rather
unexpectedly, the SUSY  provides a unif\/ied formulation of
  Laughlin and Moore--Read   states.

\section[SUSY Chern-Simons effective field theory]{SUSY Chern--Simons ef\/fective f\/ield theory}

In this section, we explore a f\/ield theoretical description for
the SUSY quantum Hall ef\/fect. First, we provide one-particle
Lagrange formalism which is  complementary to the Hamilton
formalism developed above.
 Next, we construct a Chern--Simons f\/ield theory for  the present   SUSY many-body problem.

\subsection[One-particle Lagrange formalism {[46]}]{One-particle Lagrange formalism \cite{PRD74045026}}

The one-particle  Lagrangian subject to the surface of  the
supersphere is given by
\begin{gather}
L=\frac{M}{2}(\dot{x}_a^2+C_{\alpha\beta}\dot{\theta}_{\alpha}\dot{\theta}_{\beta})
+\dot{x}_a A_a+\dot{\theta}_{\alpha}A_{\alpha}-V,
\label{oneparticlelagsupers}
\end{gather}
where $x_a$ and $\theta_{\alpha}$ satisfy the constraint
(\ref{r2supersphere}). The  covariant angular momenta
corresponding to~(\ref{lambdas})~are
\begin{gather}
\Lambda_a=M\epsilon_{abc}x_b\dot{x}_c+i\frac{M}{2}\theta_{\alpha}(\sigma_a
C)_{\alpha\beta}\dot{\theta}_{\beta},
\nonumber\\
\Lambda_{\alpha}=i\frac{M}{2}x_a(\sigma_a)_{\beta\alpha}\dot{\theta}_{\beta}
-i\frac{M}{4}\theta_{\beta}(\sigma_a)_{\beta\alpha}\dot{x}_a.\label{covariantangmomentaxtheta}
\end{gather}
It is straightforward to   conf\/irm the orthogonality
(\ref{SUSYorthogonalityLandx})  with this expression. By
introducing the Lagrange multiplier $\lambda$, the equations of
motion are derived as
\begin{gather}
M\ddot{x}_a=F_{ab}\dot{x}_b-F_{a\alpha}\dot{\theta}_{\alpha}+E_a+\lambda x_a,\nonumber\\
M\ddot{\theta}_{\alpha}=C_{\alpha\beta}(F_{a\beta}
\dot{x}_a+F_{\beta\gamma}
\dot{\theta}_{\gamma})+E_{\alpha}+\lambda
\theta_{\alpha}.\label{gatherofmotionsxtheta}
\end{gather}
From these, the Lagrangian multiplier is obtained as
\begin{gather*}
\lambda=-M(\dot{x}_a^2+C_{\alpha\beta}\dot{\theta}_{\alpha}\dot{\theta}_{\beta})-(E_a
x_a+C_{\alpha\beta}E_{\alpha}\theta_{\beta}).
\end{gather*}
Equations~(\ref{gatherofmotionsxtheta})  suggest that the super
drift motion of the particle:
\begin{gather}
E_a \dot{x}_a+C_{\alpha\beta}E_{\alpha}\dot{\theta}_{\beta} =
M(\dot{x}_a\ddot{x}_a+C_{\alpha\beta}\dot{\theta}_{\alpha}\ddot{\theta}_{\beta}).
\label{prehallortho}
\end{gather}
In the LLL limit $(M \rightarrow 0)$, the right-hand-side of
(\ref{prehallortho}) becomes zero and the electric f\/ields are
orthogonal to the SUSY Hall currents as previously discussed
(\ref{totalorthogEand}). When the electric f\/ields are
 turned of\/f, the velocity and the acceleration becomes orthogonal; this represents the circular motion around the center-of-mass coordinates.
In the LLL, the one-particle Lagrangian
(\ref{oneparticlelagsupers}) is reduced to
\begin{gather*}
L_{LLL}= \dot{x}_a A_a+\dot{\theta}_{\alpha}A_{\alpha}-V.
\end{gather*}
Since the variation of the  SUSY Hopf spinor provides the gauge
f\/ields  (\ref{SUSYBerryphaseform}),
 the gauge interaction term is simply represented as
\begin{gather*}
\dot{x}_a
A_a+\dot{\theta}_{\alpha}A_{\alpha}=-iI\psi^{\ddagger}\frac{d}{dt}\psi.
\end{gather*}
It is quite simple to see the realization of the SUSY NCG with use
of  the LLL Lagrangian.
 Regarding the Hopf spinor as  fundamental variables,  the canonical  momentum to $\psi$ is given by
\begin{gather*}
\pi=\partial L_{LLL}/\partial \dot{\psi}=-iI\psi^{\ddagger}.
\end{gather*}
The canonical quantization condition between $\psi$ and $\pi$
induces the relation:
\begin{gather*}
[\psi,\psi^{\ddagger}]_{\pm}=-\frac{1}{I},
\end{gather*}
where $+$ denotes the commutator used for Grassmann even-even,
even-odd  and odd-even components of $\psi$ and $\psi^{\ddagger}$,
while  $-$ denotes the anticommutator used for Grassmann odd-odd
case. Thus, we reproduce the results of
equation~(\ref{effectivcepsiinLLL});  complex variables are
equivalent to  derivatives in  LLL.

In the planar limit $x_3\approx R$, the one-particle Lagrangian is
reduced to
\begin{gather*}
L=\frac{M}{2}(\dot{x}_i^2+C_{\alpha\beta}\dot{\theta}_{\alpha}\dot{\theta}_{\beta})-\frac{B}{2}\epsilon_{ij}\dot{x}_i
x_j-i\frac{B}{2}(\sigma_1)_{\alpha\beta}\dot{\theta}_{\alpha}\theta_{\beta}.
\end{gather*}
  The canonical momenta $p_i=\frac{\partial}{\partial \dot{x}_i}L$, $p_{\alpha}= \frac{\partial}{\partial \dot{\theta}_{\alpha}}L $  are calculated as
\begin{gather}\label{relationpandx}
p_i=M\dot{x}_i -\frac{B}{2}\epsilon_{ij}x_j,\qquad p_{\alpha} =  M
C_{\alpha\beta}\dot{\theta}_{\beta}
-i\frac{B}{2}(\sigma_1)_{\alpha\beta}\theta_{\beta},
\end{gather}
and the Hamiltonian is obtained as
\begin{gather*}
H=\frac{1}{2M}\left(p_i+\frac{B}{2}\epsilon_{ij}x_j\right)^2
+\frac{1}{2M}C_{\alpha\beta}\left(p_{\alpha}+i\frac{B}{2}(\sigma_1\theta)_{\alpha}\right)
\left(p_{\beta}+i\frac{B}{2}(\sigma_1\theta)_{\beta}\right).
\end{gather*}
Imposing the canonical quantization conditions,
$[x_i,p_j]=i\delta_{ij}$ and
$\{\theta_{\alpha},p_{\beta}\}=i\delta_{\alpha\beta}$, it is
straightforward to derive
 the quantum mechanical Hamiltonian (\ref{planarSUSYHam}).
The relation (\ref{relationpandx}) suggests that, in the LLL
limit, the momenta are reduced to the coordinates; $p_i\rightarrow
-{B}/{2}\epsilon_{ij}x_j$  and $p_{\alpha}\rightarrow
-i{B}/{2}(\sigma_1)_{\alpha\beta}\theta_{\beta}$, and then $x_i$
and $\theta_{\alpha}$ satisfy the SUSY noncommutative algebra
(\ref{planarNC}).

\subsection[Charge-flux duality {[45]}]{Charge-f\/lux duality \cite{IJMPB52675}}

\begin{figure}[t]
  \centerline{\includegraphics[height=25mm]{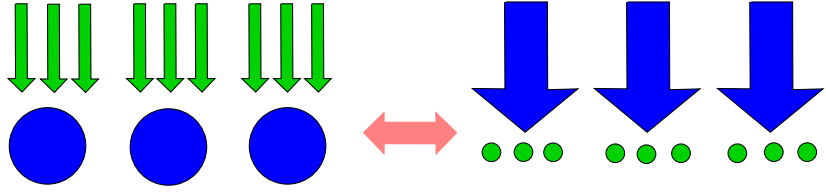}}

    \caption{The charge-f\/lux duality in the case of  $m=3$.  The left f\/igure represents the original particle-f\/lux
 system, where 3-CS f\/luxes are attached to each particle.
Meanwhile, in  the right f\/igure, 3-particles are ``attached'' to
one CS f\/lux.
The roles of particles (charges) and f\/luxes are interchanged in
the left and right f\/igures. Then, this transformation is called
charge-f\/lux duality.  }
  \label{dualityCS.fig}
\end{figure}

It is well known that the Chern--Simons f\/ield theory nicely
describes the low energy dynamics of QHE \cite{PRL8862}. Since the
Chern--Simons coupling induces the statistical transformation
specif\/ic to  3D space-time, the Chern--Simons theory plays  a
crucial role for the f\/ield theoretical description of anyons in
QHE.
In 3D particle-magnetic f\/lux  system, there is another  important concept known as the charge-f\/lux duality. 
 The charge-f\/lux duality is referred to the interchangeability of
  the matter current $J_a$ and the f\/ield strength $F_{ab}$ $(a,b=1,2,3)$.
(Here,  Wick-rotated  3D space-time $\mathbb{R}^3$ is considered.)
 Thanks to the existence of the 3-rank antisymmetric tensor, $\epsilon_{abc}$, in 3D, 2-rank antisymmetric tensor is transferred to  vector:
\begin{gather*}
F_a\equiv \tfrac{1}{2}\epsilon_{abc}F_{bc},
\end{gather*}
and hence there is one-to-one correspondence between $J_a$ and
$F_{ab}$. The charge conservation law  $\partial_aJ_a=0$ is also
consistently transfered to  the Bianchi identity $\partial_aF_a=0$
in the dual picture.
The CS  theory also provides  an appropriate f\/ield theoretical
framework to realize the charge-f\/lux duality. The CS Lagrangian
coupled to the matter current is given by
\begin{gather}
\mathcal{L}_{CS}= A_a J_a+\frac{1}{4m\pi} A_a F_a,
\label{origisimpleCSmat}
\end{gather}
where $1/m$ represents the CS coupling (that corresponds to the
f\/illing factor in
 QHE).
The equation of motion for $A_3$ is
\begin{gather*}
m\rho= \rho_{\Phi},
\end{gather*}
where $\rho$ represents the particle density $J_3$, and
$\rho_{\Phi}$ represents the CS magnetic f\/lux density $B/2\pi$.
This relation manifests that  the $m$-CS f\/luxes are attached  to
each particle.
 Since   the currents in the original system  correspond to  dual f\/ield strengths,  the  Lagrangian (\ref{origisimpleCSmat}) may be  rewritten~as
\begin{gather*}
\mathcal{L}_{CS} =A_a \tilde{F}_a+\frac{1}{4m\pi} A_a F_a,
\end{gather*}
where $\tilde{F}_a$ denote the dual f\/ield strengths. Integrating
out the original CS f\/ields, we obtain the dual Lagrangian
expressed by the dual CS f\/ields,
\begin{gather*}
\tilde{\mathcal{L}}_{CS}=-{m\pi}\tilde{A}_a\tilde{F}_a.
\end{gather*}
The CS coupling in the dual CS Lagrangian is inverse to that  in
the original CS Lagrangian; the strong CS coupling region in  the
original system corresponds to the weak coupling region in the
dual system, and vice versa\footnote{In this sense, the
charge-f\/lux duality corresponds to the $S$-dual transformation
of the Chern--Simons coupling in the modern string theory
language.} (see Fig.~\ref{dualityCS.fig} also). The charge-f\/lux
duality is a very important concept for the study of  topological
objects, since the existence of the duality permits us to switch
to the dual  description where topological objects arise as
fundamental excitations.

\subsection[The SUSY Chern-Simons description {[46]}]{The SUSY Chern--Simons description \cite{PRD74045026}}

We show how the charge-f\/lux duality   is naturally   generalized
in
  the SUSY  QHE.
In the Euclidean super space-time $\mathbb{R}^{3|2}$, there exist
the super matter currents $J_a$, $J_{\alpha}$ and the 2-rank super
f\/ield strengths $F_{ab}$, $F_{a\alpha}$, $F_{\alpha\beta}$.
The super f\/ield strengths are related to each other by the  SUSY
transformations generated by $Q=L_{\alpha}\xi_{\alpha}$:
\begin{gather*}
\delta_{\xi}F_{ab}=-\tfrac{1}{2}F_{a\alpha}(C\sigma_b
\xi)_{\alpha}+\tfrac{1}{2}(C\sigma_a\xi)_{\alpha},\qquad
\delta_{\xi}F_{a\alpha}=\tfrac{1}{2}F_{ab}(\sigma_b\xi)_{\alpha}+\tfrac{1}{2}F_{\alpha\beta}(C\sigma_a\xi)_{\beta},\\
\delta_{\xi}F_{\alpha\beta}=-\tfrac{1}{2}F_{a\alpha}(\sigma_b\xi)_{\beta}-\tfrac{1}{2}F_{a\beta}(\sigma_a\xi)_{\alpha}.
\end{gather*}
Since   the number of components of the super-vector currents
($=5$) and that of  2-rank super tensor f\/ield strengths ($=12$)
do not match, one may suspect  whether   the charge-f\/lux duality
exists in the SUSY case. However,  $\bold{12}$-dimensional 2-rank
tensors  are irreducibly decomposed to $\bold{5}\oplus \bold{7}$.
The $\bold{5}$-dimensional f\/ield strengths, which we call the
super-vector f\/ield strengths, are explicitly constructed as
\begin{gather*}
F_a\equiv
\tfrac{1}{2}\epsilon_{abc}F_{bc}+i\tfrac{1}{4}(C\sigma_a)_{\alpha\beta}F_{\alpha\beta},\qquad
 F_{\alpha}\equiv -i\tfrac{1}{2}(C\sigma_a)_{\alpha\beta}F_{a\beta}.
\end{gather*}
Indeed, under the $OSp(1|2)$ SUSY transformation, they form a
multiplet:
\begin{gather*}
\delta_{\xi}F_a=\tfrac{1}{2}F_{\alpha}(\sigma_a
\xi)_{\alpha},\qquad
\delta_{\xi}F_{\alpha}=\tfrac{1}{2}F_a(C\sigma_a\xi)_{\alpha}.
\end{gather*}
With the super-vector f\/ield strengths, it is possible to develop
the charge-f\/lux duality even in the SUSY case. There exists
one-to-one correspondence between the matter currents and the
 super-vector f\/ield strengths,
\begin{gather*}
J_a  \ \leftrightarrow \  F_a,\qquad J_{\alpha}  \ \leftrightarrow
\ F_{\alpha},
\end{gather*}
and  the charge conservation is consistently transfered to the
Bianchi identity again: $\partial_a
J_a+\partial_{\alpha}J_{\alpha}=0~ \leftrightarrow~ \partial_a
F_a+
\partial_{\alpha}F_{\alpha}=0.$
Taking the inner product between $(A_a,A_{\alpha})$ and
$(F_a,F_{\alpha})$, our  SUSY  Chern--Simons
Lagrangian\footnote{There are various types of SUSY CS theories,
for instance   \cite{Nissimov,PRD451992,hep-th/9612031}. Here, we
develop a new type of SUSY CS theory that is def\/ined on
supermanifold. The matrix version of (\ref{ourSUSYCSLag}) plays a
crucial role for  realization of  fuzzy supersphere in supermatrix
model \cite{hep-th/0311005}. } is constructed   as
\begin{gather}
\mathcal{L}_{sCS}=F_aA_a+F_{\alpha}A_{\alpha}
=\epsilon_{abc}A_a\partial_bA_c-i(C\sigma_a)_{\alpha\beta}
A_{\alpha}\partial_aA_{\beta}+2i(C\sigma_a)_{\alpha\beta}A_{\alpha}\partial_\beta
A_a. \label{ourSUSYCSLag}
\end{gather}
The SUSY Chern--Simons Lagrangian (\ref{ourSUSYCSLag}) possesses
the apparent $OSp(1|2)$ global symmetry and  the $U(1)$ gauge
invariance up to total derivatives:
\begin{gather*}
\delta \mathcal{L}_{sCS}=\partial_a{ (\Lambda F_a )}+\partial_{\alpha}(\Lambda F_{\alpha})\nonumber\\
\phantom{\delta \mathcal{L}_{sCS}}{}
=\tfrac{1}{2}\partial_a(\Lambda\epsilon_{abc}  F_{bc})
-i\tfrac{1}{2}
  \partial_{\alpha}(\Lambda(C\sigma_a)_{\alpha\beta}F_{a\beta})  +i\tfrac{1}{4}\partial_a( \Lambda (C\sigma_a)_{\alpha\beta}F_{\alpha\beta} ),
\end{gather*}
where $\Lambda$ is the $U(1)$ gauge parameter. It is also possible
to show  that our SUSY CS Lagrangian possesses topological
properties analogous to
  the  original Chern--Simons theory;
it exhibits SUSY linking number,    SUSY topological mass
generations  and etc \cite{PRD74045026}. With this  SUSY   CS
term,
 the Chern--Simons--Landau--Ginzburg (CSLG) Lagrangian is constructed as
\begin{gather}
\mathcal{L}_{CSLG}=A_a
J_a+A_{\alpha}J_{\alpha}+\frac{1}{4m\pi}(F_aA_a+F_{\alpha}A_{\alpha})
+\cdots, \label{SUSYCSLG}
\end{gather}
where $\cdots$ includes the  kinetic term of matter f\/ield, the
Coulomb potential energy and etc. Substituting  the dual CS
f\/ields for the matter currents, the SUSY CS Lagrangian with
interaction term is expressed as
\begin{gather}
\mathcal{L}= \mathcal{L}_I+\mathcal{L}_{sCS} = (A_a\tilde{F}_a+
A_{\alpha}\tilde{F}_{\alpha})+\frac{1}{4m\pi}(A_a
F_a+A_{\alpha}F_{\alpha}). \label{SUSYCSPURE}
\end{gather}
Taking  advantage of the duality,  the dual  CSLG Lagrangian is
systematically derived~\cite{PRD74045026}. For instance,
integrating out the original CS f\/ields $(A_a,A_{\alpha})$ in
equation~(\ref{SUSYCSPURE}), we obtain the dual CS Lagrangian:
\begin{gather*}
\tilde{\mathcal{L}}_{sCS}=-{m\pi}(\tilde{A}_a\tilde{F}_a+\tilde{A}_{\alpha}\tilde{F}_{\alpha})\nonumber\\
\phantom{\tilde{\mathcal{L}}_{sCS}}{}
=-\frac{m\pi}{2}\left(\epsilon_{abc}\tilde{A}_a\tilde{F}_{bc}-
i(C\sigma_a)_{\alpha\beta}\tilde{A}_{\alpha}\tilde{F}_{a\beta}
+\tfrac{i}{2}(C\sigma_a)_{\alpha\beta}\tilde{A}_a\tilde{F}_{\alpha\beta}\right).
\end{gather*}
 The dual CS Lagrangian is  identical  to the original SUSY CS Lagrangian (\ref{SUSYCSLG}) except for  the inverse CS coupling, as found in the original bosonic case.
In a low energy limit, the dual CSLG Lagrangian  takes the  form
\begin{gather*}
L_{\rm ef\/f}=2\pi\sum_p(\dot{x}^p_i\tilde{A}_i
+\dot{\theta}_{\alpha}^p\tilde{A}_{\alpha})-V+\tilde{\mathcal{L}}_{sCS},
\end{gather*}
where $x_i^p$ and $\theta^p_{\alpha}$ denote the position of the
$p$-th vortex on the superplane, and $V$ denotes the Coulomb
potential term. From $L_{\rm ef\/f}$, the
 equation of motion for  vortex is derived as
\begin{gather}\label{orthotildeFandE}
2\pi(-\tilde{F}_{ij}\dot{x}^p_j+\tilde{F}_{i\alpha}\dot{\theta}^p_{\alpha})=E_i,\qquad
2\pi(\tilde{F}_{i\alpha}\dot{x}^p_i+\tilde{F}_{\alpha\beta}\dot{\theta}^p_{\beta})=C_{\alpha\beta}E_{\beta}.
\end{gather}
Equations~(\ref{orthotildeFandE}) suggest that the vortex moves
perpendicularly to the direction of the applied super electric
f\/ields:
\begin{gather*}
E_i \dot{x}^p_i+C_{\alpha\beta}E_{\alpha}\dot{\theta}^p_{\beta}=0,
\end{gather*}
which   manifests the Hall orthogonality in the SUSY sense.

\section{Summary and discussion}

We overviewed the developments of the SUSY QHE. It was  shown that
the framework of  QHE was naturally supersymmetrized based on the
SUSY Hopf map. In the construction of the SUSY  QHE we have
encountered many exotic mathematical and physical ideas. The SUSY
Hopf f\/ibration was  crucial  in construction of the spherical
SUSY QHE. In the LLL limit,  the fuzzy supersphere naturally
emerges. In the planar SUSY QHE, we explored the SUSY Landau
problem, and found the existence of $\mathcal{N}=2$ SUSY. (The
existence of the $\mathcal{N}=2$ SUSY may be a~generic feature of
SUSY planar Landau models \cite{hep-th/0510019,hep-th/0612300}.)
With appropriate interpretation of the Grassmann  quantity, we
have shown that the SUSY Laughlin wavefunction contains   the
original Laughlin
 and the Moore--Read states as its f\/irst and last component wavefunctions.
A~SUSY CS f\/ield theory  is also  developed as the appropriate
ef\/fective f\/ield theory  for the  SUSY QHE. The newly derived
Chern--Simons theory is invariant under the  global $OSp(1|2)$
and local $U(1)$ transformations, and shares topological features
with the original CS theory. The  Hall orthogonality  and the
charge-f\/lux duality
 are consistently generalized in the  SUSY framework.

 However, there still remain many issues to be addressed within the formulation of the SUSY QHE, such as edge excitations, hydrodynamic description and relations to integrable systems.
Among them, one of the most important issues  is to explore
applications to  real condensed matter physics. As we have seen,
the SUSY brings a unif\/ied picture of  the original Laughlin and
the Moore--Read states. It would be  worthwhile to speculate  what
insights  such    unif\/ication could yield  to the original QHE.
Though the SUSY QHE  provides a concrete physical realization of
the non-anti-commutative geometry, our set-up is still restricted
to  low dimensions. It is quite tempting to extend our SUSY
formulation to  higher dimensions. The construction of higher
dimensional SUSY QHE  may be benef\/icial to the understanding of
higher dimensional fuzzy super geometries, in particular classical
solutions  of  supermatrix models.  Besides, as reported
in~\cite{cond-mat/0211679,cond-mat/0401224},
 QHE contains  mathematical structures similar to  the twistor theory.
It is also  interesting to exploit relations between the SUSY QHE
and supertwistor theory. Further developments of QHE may bring
fruitful consequences in a wide realm of  modern physics.

\subsection*{Acknowledgements}

I would like to thank  Yusuke Kimura for  collaborations in  early
stage of this work.

\pdfbookmark[1]{References}{ref}
\LastPageEnding

\end{document}